# Co-current rotation of the bulk ions due to the ion orbit loss at the edge of a tokamak plasma

**Chengkang Pan[1], Shaojie Wang[2], Jing Ou[1]**

[1]Institute of Plasma Physics, Chinese Academy of Science, Hefei, 230031, China
[2]Department of Modern Physics, University of Science and Technology of China, Hefei, 230026, China

E-mail: ckpan@ipp.ac.cn

**ABSTRACT**

Flux-surface-averaged momentum loss and parallel rotation of the bulk ions at the edge of a tokamak plasma due to the ion orbit loss are calculated by computing the minimum loss energy of both the trapped and the passing thermal ions. The flux-surface-averaged parallel rotation of the bulk ions is in the co-current direction. The peak of the co-current rotation speed locates inside the last closed flux surface due to the orbit loss of the co-current thermal ions at the very edge of a tokamak plasma. The peaking position moves inward when the ion temperature increases.

**I. Introduction**

Plasma rotation and radial electric field (REF) play important roles in the transition from low (L) to high (H) confinement modes [1-3] as well as the formation of internal transport barriers (ITBs) in tokamaks [4, 5]. Plasma rotation is beneficial for improving confinement and avoiding critical instabilities [6] and is mainly driven by the neutral beam injection (NBI) in the present tokamaks. However, in a reactor-grade tokamak such as ITER, the NBI can not drive strong rotation, since the power density of the planed NBI is lower than that of the present tokamaks. It has been observed that the tokamak plasmas rotate even in the absence of external momentum injection, which is known as the intrinsic rotation [7-10].

There is experimental evidence indicating that the momentum source (or sink) causing the intrinsic rotation is located in the tokamak edge [11-13]. The ion orbit loss has long been thought to play an important role in generating the REF and the plasma rotation at the edge of a tokamak plasma [14-30].

Intrinsic rotation at the edge of a tokamak plasma due to the ion orbit loss has been studied theoretically by many authors [22, 26, 27, 30]; these theoretical works based on the ion orbit loss model predict that the peak of the co-current rotation speed locates at the last closed flux surface (LCFS). However, the experimental observations [12, 31] and the edge gyrokinetic simulations [21] indicate that the peak of the co-current rotation speed locates inside the LCFS. Therefore, it is of interest to investigate the reason of this important discrepancy.

In this paper, we will show that the calculation of the minimum loss energy of the counter-current passing ions in Refs. [24 and 30] was incorrect and the orbit loss of the co-current ions was neglected in Refs. [22 and 26]. Having correctly calculated the ion orbit loss, we found that the peak of the co-current rotation speed locates inside the LCFS. This is due to the orbit loss of the co-current thermal ions at the very edge of the tokamak plasma.

The remaining part of this paper is organized as follows. In Sec. II, the ion orbit loss model is presented. In Sec. III, the minimum loss energy and the loss fractions are calculated. In Sec. IV, the flux-surface-averaged parallel rotation of the bulk ions due to the ion orbit loss is calculated. The conclusions are presented in Sec. V.

**II. Ion orbit loss model**

### A. Guiding-center orbit

We start from the ion guiding-center orbit in a tokamak. The guiding-center orbit of an ion in an axisymmetry tokamak is determined by the three independent constants of motion, i.e., the magnetic moment $\mu$, the total energy $E$, and the canonical toroidal angular momentum $P_\alpha$ [23, 32, 33]:

$$\mu = \frac{m_s V_\perp^2}{2B}, \tag{1}$$

$$E = \frac{1}{2}m_s(V_\parallel^2 + V_\perp^2) + e_s\Phi, \tag{2}$$

$$P_\alpha = \psi + IV_\parallel/\Omega. \tag{3}$$

Here $m_s$, $e_s$, $V_\parallel$ and $V_\perp$ are the ion's mass, electric charge, and velocity components parallel and perpendicular to the magnetic field, respectively. $B$ and $\Phi$ are the equilibrium magnetic field and the electrostatic potential. $\psi$ is the poloidal magnetic flux and $\Omega = e_s B/m_s$ is the gyrofrequency. $I = RB_t$ is a constant, with $R$ the major radius and $B_t$ the toroidal magnetic field. These three constants of motion will be used to determine which particle may be lost out of the LCFS due to the magnetic drift. The effect of the REF shall be ignored in this paper, which should not modify the results qualitatively [26].

A large-aspect-ratio tokamak with circular cross section is considered for simplicity. We assume that the plasma current flows in the counter-clockwise direction looking down on the tokamak and the toroidal magnetic field is in the opposite direction. Specifying a uniform current density, one can obtain the poloidal magnetic flux

$$\psi = \frac{1}{2}\left(\frac{\mu_0 I_p}{2\pi a^2}\right)a^2 \bar{R}\rho^2, \tag{4}$$

where $a$ is the minor radius of the plasma, $I_p$ and $\bar{R}$ are the plasma current and the major radius at the magnetic axis, and $\rho = r/a$ is the normalized minor radius. The poloidal magnetic flux can also be written in the form

$$\psi = \psi(R, Z), \tag{5}$$

where $Z$ is the vertical coordinate and $Z = 0$ corresponds to the midplane of the tokamak.

The ion orbit constraints, i.e., Eq. (1)-(3), can also be written in the $(R, \psi)$ plane,

$$(\psi - P)^2 = \left(\frac{I}{\Omega}\right)^2 \left[2E/m_s - V_0^2(1-\zeta_0^2)B/B_0\right], \tag{6}$$

where $V_0$ is the initial speed of the ion and $\zeta_0 = -\vec{V}_0 \cdot \vec{B}/V_0 B$. $\zeta_0 > 0$ denotes that the ion moves in the co-current direction initially. $B_0$ is the magnetic field at the starting position.

Typical ion guiding-center orbits in the $(R,\psi)$ plane are shown in Figure 1. The ion guiding-center orbit will have two intersections with the midplane of the tokamak except the very barely trapped ion. These two intersections correspond to the innermost flux surface, $\psi_{min}$, and the outermost flux surface, $\psi_{max}$, that the ion can reach. We can see from Figure 1 that the two intersections of the trapped ion are at the outboard midplane. And for the passing ion, one of the intersections is at the outboard midplane and the other is at the inboard midplane.

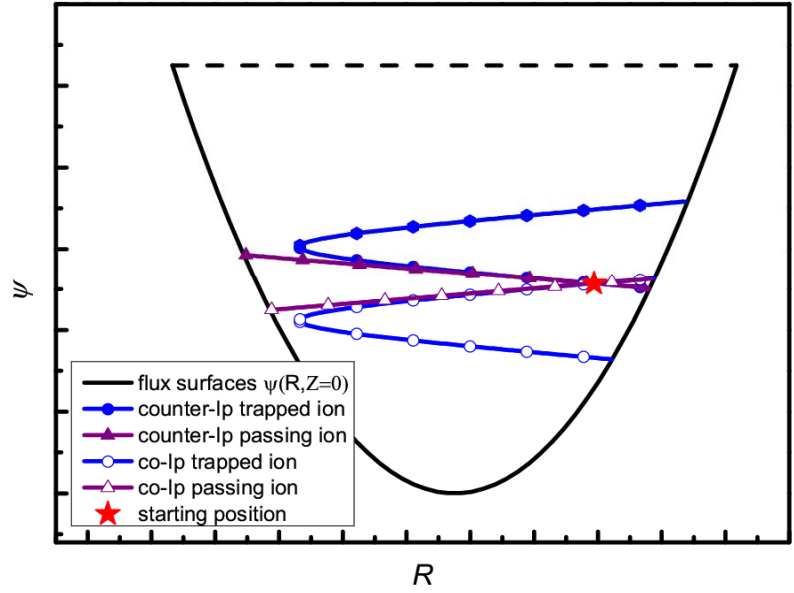

Figure 1. Typical ion guiding-center orbits in the $(R,\psi)$ plane.

**B. Types of the loss orbit and determination of the minimum loss energy**

The ion which starts from the position $(\psi_0,\theta_0)$ will be lost out of the LCFS at the position $(\psi_a,\theta_a)$ if $\psi_{max}>\psi_a$, where $\theta$ is the poloidal angle, the subscript '0' denotes the corresponding value at the starting position and the subscript 'a' denotes the corresponding value at the LCFS respectively.

The co-current ions starting from the outboard midplane, $(\psi_0,\theta_0=0)$, and the counter-current passing ions starting from the inboard midplane, $(\psi_0,\theta_0=\pi)$, will not be lost due to $\psi_0=\psi_{max}\leq\psi_a$ and their radially inward drift. The ions locating initially on the LCFS, i.e., $\psi_0=\psi_a$, will be lost out of the LCFS except the co-current ions starting from the outboard midplane and the counter-current ions starting from the inboard midplane.

Generally there are three types of loss orbit, which are shown in Figure 2. For the lost trapped ion, it reaches the outmost flux surface, $\psi_{max}$, at the outboard midplane. For the lost counter-current passing ion, it reaches the outmost flux surface, $\psi_{max}$, at the inboard midplane. For the lost co-current passing ion, it reaches the outmost flux surface, $\psi_{max}$, at the outboard midplane. Note that the three types of loss orbit shown in Figure 2 are the critical cases, which correspond to the particles just touches the LCFS.

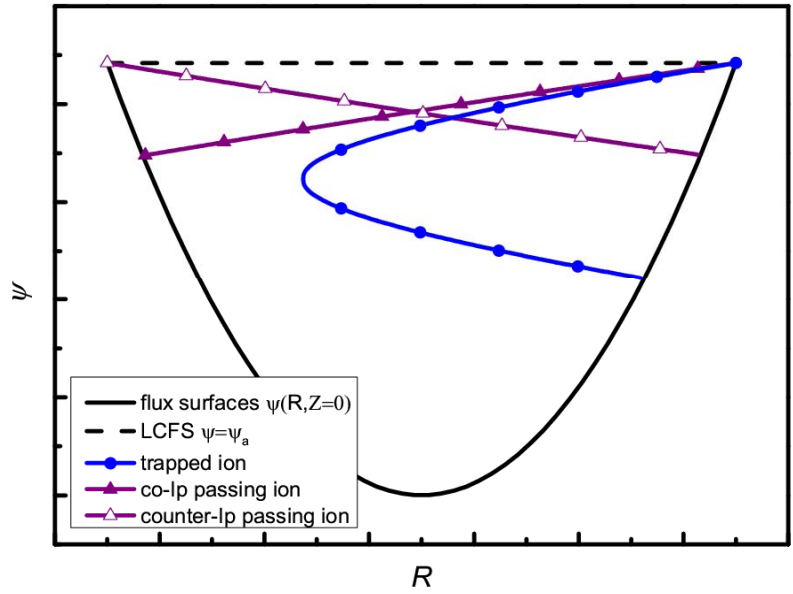

Figure 2. Three types of loss orbit, with $\psi_{max} = \psi_a$.

An ion drifting radially outward can be lost out of the LCFS if its kinetic energy is greater than a critical value, the minimum loss energy, $E_{min} = m_s V_{0,min}^2 / 2$. The losses of particle, energy and momentum due to the ion orbit loss can be calculated based on the minimum loss energy. Correct determination of the minimum loss energy is important for studying the ion orbit loss physics.

The minimum loss energy of an ion starting from the position $(\psi_0, \theta_0)$ with the initial $\zeta_0$ is the energy for this ion to reach the $\psi_{max} = \psi_a$ flux surface at the poloidal angle $\theta_a$. Three types of ion loss orbit with $\psi_{max} = \psi_a$ are shown in Figure 2. For the barely lost trapped ion, it reaches the

$\psi_{\max} = \psi_a$ flux surface at the outboard midplane, $\theta_a = 0$. For the barely lost counter-current passing ion, it reaches the $\psi_{\max} = \psi_a$ flux surface at the inboard midplane, $\theta_a = \pi$. For the barely lost co-current passing ion, it reaches the $\psi_{\max} = \psi_a$ flux surface at the outboard midplane, $\theta_0 = 0$. Based on this analysis of the three types of barely lost orbit, we shall calculate the minimum loss energy for these three types in the following.

The equations to determine the minimum loss energy of an ion initially locating at $(\psi_0, \theta_0)$ with a given $\zeta_0$ are [24, 28-30]

$$\psi_a + IV_{//a}/\Omega_a = \psi_0 + IV_{//0}/\Omega_0, \tag{7}$$

$$\frac{1}{2}m_s(V_{//a}^2 + V_{\perp a}^2) = \frac{1}{2}m_s(V_{//0}^2 + V_{\perp 0}^2), \tag{8}$$

$$\frac{m_s V_{\perp a}^2}{2B_a} = \frac{m_s V_{\perp 0}^2}{2B_0}, \tag{9}$$

where the subscript '0' denotes the corresponding value at the starting position and the subscript 'a' denotes the corresponding value at the position $(\psi_{\max} = \psi_a, \theta_a)$ that has been discussed above.

From Eq. (8) and Eq. (9), we can obtain

$$V_{//a} = \pm V_{0,\min}\left[1 - \frac{B_a}{B_0}(1 - \zeta_0^2)\right]^{1/2}, \tag{10}$$

The sign of $V_{//a}$ should be consistent with the type of the ion loss orbit. For the barely lost passing ion, the sign of $V_{//a}$ should be the same as that of $V_{//0}$. For the barely lost trapped ion, the sign of $V_{//a}$ is positive. For the barely lost trapped ion with $\zeta_0 > 0$, the sign of $V_{//a}$ is the same as that of $V_{//0}$. For the barely lost trapped ion with $\zeta_0 < 0$, the sign of $V_{//a}$ is opposite to that of $V_{//0}$.

Substituting Eq. (10) into Eq. (7), one finds the minimum loss speed of the ion

$$V_{0,\min} = \frac{\psi_a - \psi_0}{I/\Omega_0}\left\{\zeta_0 \pm \frac{B_0}{B_a}\left[1 - \frac{B_a}{B_0}(1 - \zeta_0^2)\right]^{1/2}\right\}^{-1}. \tag{11}$$

The sign of the second term in Eq. (11) should be consistent with that discussed above and the value of $B_a$ should be evaluated at the position $(\psi_{\max} = \psi_a, \theta_a)$. From Eq. (11), we can see that the minimum loss speed depends on the initial pitch angle and the starting position of the ion. Combining Eq. (11) with (4), we can calculate the minimum loss energy, $E_{\min} = m_s V_{0,\min}^2/2$, of an ion with the initial values $(\psi_0, \theta_0, \zeta_0)$.

**III. Loss fractions due to the ion orbit loss**

After the minimum loss energy is determined, the particle and the momentum loss fractions can be calculated. An initial Maxwellian distribution with the local ion temperature, $T_i$, is assumed. The corresponding cumulative loss fractions are defined following Stacey's definitions [24, 28-30]:

$$P(\rho,\theta_0) = \int_{-1}^{1} P_{\zeta_0}(\rho,\theta_0) d\zeta_0 = \int_{-1}^{1} \frac{\int_{V_{0,\min}}^{\infty} V_0^2 f(V_0) dV_0}{2\int_0^{\infty} V_0^2 f(V_0) dV_0} d\zeta_0, \qquad (12)$$

$$M(\rho,\theta_0) = \int_{-1}^{1} M_{\zeta_0}(\rho,\theta_0) d\zeta_0 = \int_{-1}^{1} \frac{\int_{V_{0,\min}}^{\infty} (m_s V_0 \zeta_0) V_0^2 f(V_0) dV_0}{2\int_0^{\infty} (m_s V_0) V_0^2 f(V_0) dV_0} d\zeta_0, \qquad (13)$$

where $f(V_0) = (m_s/2\pi T_i)^{3/2} \exp(-m_s V_0^2/2T_i)$, $P_{\zeta_0}$ and $M_{\zeta_0}$ are the $\zeta_0$ dependent particle and the momentum loss fractions respectively.

The model tokamak parameters are chosen as: ($B_{t0} = -1.6T, \bar{R} = 1.7m, a = 0.6m, I_p = 1.3MA$). The toroidal magnetic field is opposite to the plasma current as mentioned in Section II. We only consider the bulk ions $D^+$. The temperature profile used in the calculations is shown in Figure 3.

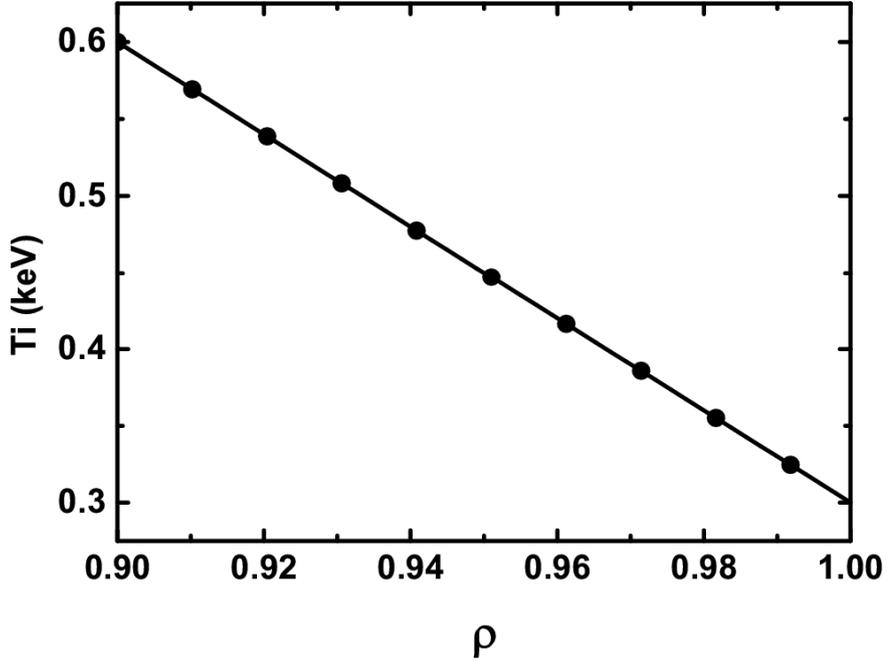

Figure 3. Profile of the ion temperature at the edge.

Firstly we calculate the minimum loss energy. The minimum loss energy of the ions starting from different poloidal positions on the $\rho = 0.98$ and the $\rho = 0.99$ flux surfaces versus $\zeta_0$ are plotted in

Figure 4(a)-(b). There are several differences between our results and Stacey's results [24, 30] as listed in the following.

(1) The minimum loss energy of the counter-current passing ions is much larger than that of the counter-current trapped ions, as is shown in Figure 4(a)-(b). The minimum loss energy of the counter-current passing ions is smaller than that of the counter-current trapped ions in Refs. [24 and 30], which is incorrect. The reason for these incorrect calculations in Refs. [24 and 30] is that $B_a$ was incorrect evaluated at the outboard midplane in their calculations for the counter-current passing ions. The point is that the barely lost counter-current passing ions reach the $\psi_{max} = \psi_a$ flux surface at the inboard midplane and the value of $B_a$ at the inboard midplane should be used. This has been discussed in Section II.

(2) The counter-current passing ions starting from the inboard midplane will not be lost, as is shown in Figure 4(a)-(b), and this is due to $\psi_0 = \psi_{max} < \psi_a$ and their radially inward drift. Note that in Refs. [24 and 30], these ions were incorrectly put into loss orbits.

(3) Also the co-current ions starting from the outboard midplane will not be lost, as is shown in Figure 4(a)-(b), and this is due to $\psi_0 = \psi_{max} < \psi_a$ and their radially inward drift. Note that in Refs. [24 and 30], these ions were also incorrectly put into loss orbits.

The incorrect calculations of the minimum loss energy will result in the incorrect loss fractions results.

The $\zeta_0$ dependent particle and momentum loss fractions, $P_{\zeta_0}$ and $M_{\zeta_0}$, of the ions starting from different poloidal positions on the $\rho = 0.98$ and the $\rho = 0.99$ flux surfaces are plotted versus $\zeta_0$ in Figure 4(c)-(f). Our $P_{\zeta_0}$ and $M_{\zeta_0}$ of the counter-current passing ions are smaller than that shown in Refs. [24 and 30]. $P_{\zeta_0}$ and $M_{\zeta_0}$ of the counter-current passing ions were overestimated in Refs. [24 and 30]. The overestimation of $M_{\zeta_0}$ of the counter-current passing ions will result in the overestimation of the co-current plasma rotation. $P_{\zeta_0}$ and $M_{\zeta_0}$ of the co-current ions increase with $\theta_0$. The co-current ions starting from the outboard midplane can not be lost as mentioned above. However, there were non-negligible loss fractions for the co-current passing ions starting from the outboard midplane in Ref. [30].

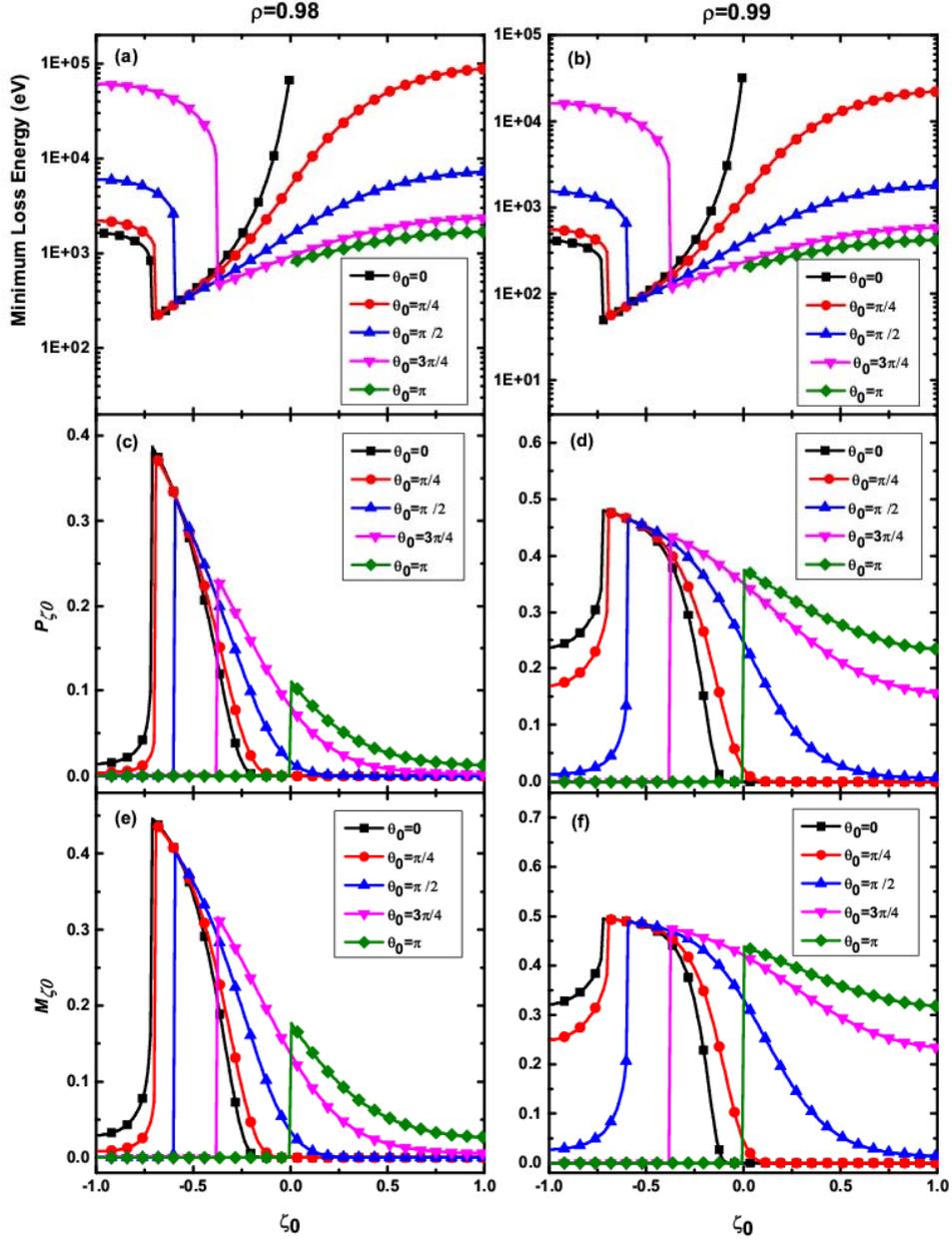

Figure 4. (a) and (b) Minimum loss energy, (c) and (d) $\zeta_0$ dependent particle loss fractions $P_{\zeta_0}$, (e) and (f) $\zeta_0$ dependent momentum loss fractions $M_{\zeta_0}$ for the ions starting from different poloidal positions on the $\rho = 0.98$ and the $\rho = 0.99$ flux surfaces.

The cumulative particle and momentum loss fractions of the ions on the $\rho = 0.98$ and the $\rho = 0.99$ flux surfaces are plotted versus $\theta_0$ in Figure 5. The positive and negative sign of the momentum loss denote the co-current and the counter-current momentum loss respectively. The cumulative particle loss fractions increase with the minor radius for all $\theta_0$. Then the flux-surface-averaged particle loss fractions increase with the minor radius, which is shown in Figure 6(b). The cumulative counter-current momentum loss fraction of the ions starting from the low field side increases with the minor radius. And the cumulative co-current momentum loss fraction of the ions starting from the high field side also increases with the minor radius. This results in the flux-surface-averaged momentum loss fraction varying non-monotonically with the minor radius and peaking inside the LCFS. This is shown in Figure 6(c). The co-current momentum loss and its increasing with the minor radius play important roles on the rotation profile caused by the ion orbit loss and this will be discussed in the next section.

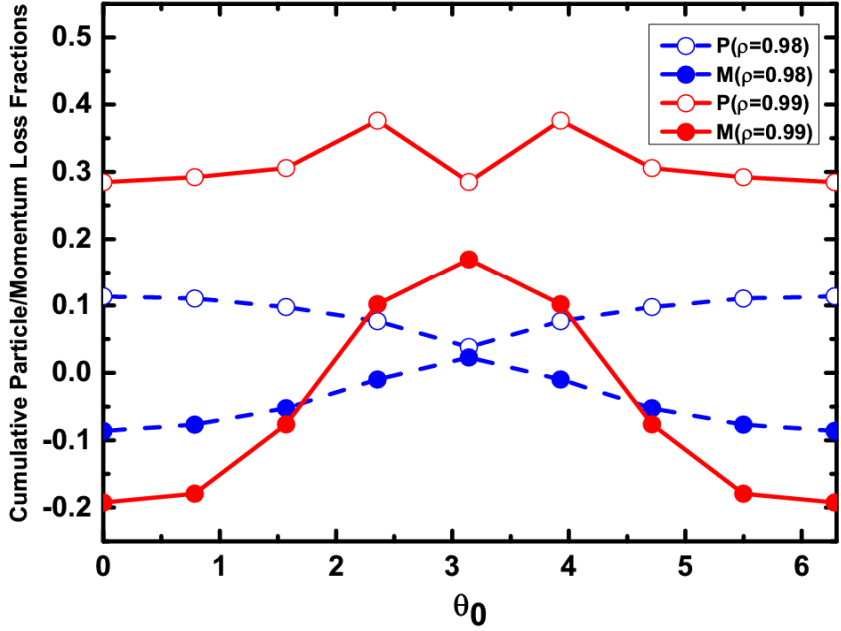

Figure 5. Cumulative particle and momentum loss fractions of the ions starting from the $\rho = 0.98$ and the $\rho = 0.99$ flux surfaces versus the starting poloidal angle $\theta_0$.

**IV. Plasma rotation due to the ion orbit loss**

From Figure 4(c)-(f), we can see that both the co-current and the counter-current ions can be lost,

but the counter-current loss is dominant. This will result in the co-current plasma rotation. The parallel rotation of the bulk ions due to the ion orbit loss can be given by [27-30]

$$V_{//}(\rho,\theta_0) = 2\pi \int_{-1}^{1} \left[ \int_{V_{0,\min}}^{\infty} (V_0 \zeta_0) V_0^2 f(V_0) dV_0 \right] d\zeta_0$$
$$= 2\pi^{-1/2} M(\rho,\theta_0) V_{Ti}$$
$$= 2\pi^{-1/2} M(\rho,\theta_0) \sqrt{2T_i/m_s}$$

(14)

The cumulative momentum loss fraction has been calculated in Section III. The flux-surface-averaged parallel rotation of the bulk ions due to the ion orbit loss on the given flux surface will be obtained by averaging Eq. (14) over the flux surface

$$<V_{//}>(\rho) = 2\pi^{-1/2} <M(\rho,\theta_0)> \sqrt{2T_i/m_s} \quad , \qquad (15)$$

where $<A(\rho,\theta_0)> = (1/2\pi)\int_0^{2\pi} A(\rho,\theta_0)(1+r/R_0 \cos\theta_0) d\theta_0$.

With the ion temperature given, the flux-surface-averaged parallel rotation of the bulk ions on a given flux surface due to the ion orbit loss can be determined. Repeating the above calculations on each flux surface at the edge of a tokamak plasma, the rotation profile of the bulk ions can be obtained.

With two different ion temperature profiles (case 1 and case 2) given in Figure 6(a), the flux-surface-averaged particle and momentum loss fractions are shown in Figure 6(b) and Figure 6(c). The case 1 has been used in Section III. The negative sign of the flux-surface-averaged momentum loss fractions denote the counter-current momentum loss. The flux-surface-averaged particle loss fraction increases with the minor radius, and the higher the ion temperature the larger the loss fraction. The flux-surface-averaged momentum loss fraction increases with the minor radius and then decreases to zero at the LCFS. The zero flux-surface-averaged momentum loss fraction on the LCFS is due to the balanced counter-current and co-current momentum loss. The non-monotonic variation of the flux-surface-averaged momentum loss fraction with the minor radius is due to the non-negligible co-current momentum loss at the very edge. The peaking position moves inward for the higher ion temperature case (case 2). The reason that the peak of the flux-surface-averaged parallel rotation speed of the bulk ions locates inside the LCFS [shown in Figure 6(d)] is clearly that the peak of the flux-surface-averaged momentum loss fraction locates inside the LCFS [shown in Figure 6(c)]. This peaking phenomenon is consistent with the experimental observations [12, 31] and the edge gyrokinetic simulation results [21] qualitatively.

The plasma rotation speed peaking inside the LCFS can be analyzed from the calculations in section III (see Figure 4). The $\rho = 0.98$ flux surface is inside the peaking position ($\rho = 0.984$). On the $\rho = 0.98$ flux surface, the momentum loss fraction of the co-current ions is much less than that of the counter-current ions, as is shown in Figure 4(e). On the $\rho = 0.99$ flux surface, the momentum loss fraction of the co-current ions is comparable with that of the counter-current ions, as is shown in Figure 4(f). However, the co-current momentum loss due to the orbit loss of the co-current ions is neglected in Refs. [22 and 26]; this is incorrect in the very edge of a tokamak plasma. The co-current momentum loss near the very edge and its increasing with the minor radius is the key to understand that the peak of the flux-surface-averaged parallel rotation speed inside the LCFS.

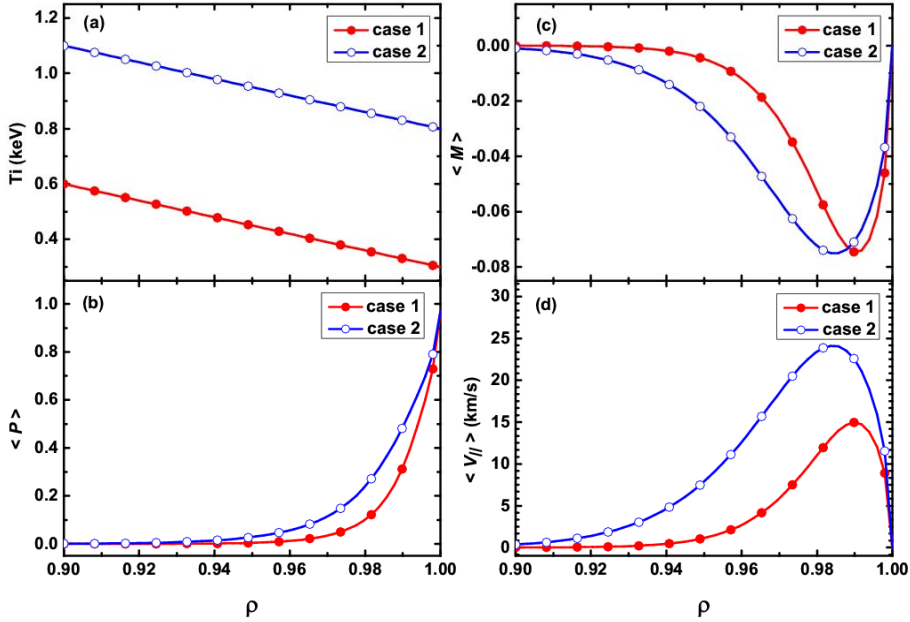

Figure 6. Profiles of (a) the ion temperature, (b) the flux-surface-averaged particle loss fraction, (c) the flux-surface-averaged momentum loss fraction, (d) the flux-surface-averaged parallel rotation of the bulk ions.

## V. Conclusions

Based on the three constants of motion, the minimum loss energy of both the trapped and the passing thermal ions at the edge of a tokamak plasma is derived. The flux-surface-averaged loss fractions of particle and momentum are calculated. The flux-surface-averaged particle loss fraction increases with the minor radius, but the flux-surface-averaged momentum loss fraction peaks inside the LCSF. The non-monotonic variation of the flux-surface-averaged momentum loss fraction with the minor radius is due to the orbit loss of the co-current thermal ions at the very edge of a tokamak plasma. The resulting co-current rotation speed of the bulk ions peaks inside the LCSF. The peaking position moves inward when the ion temperature increases. The calculated plasma rotation near the edge due to the ion orbit loss is consistent with the experimental observations and the gyrokinetic simulation results.

**ACKNOWLEDGMENTS**

This work was supported by the National Natural Science Foundation of China under Grant No. 11105182, No. 11175178, No. 11105176, and the National ITER program of China under Contract No. 2014GB113000.